\documentclass[a4paper,fleqn,usenatbib]{mnras}
\hypersetup{draft}
\usepackage{newtxtext,newtxmath}

\usepackage[T1]{fontenc}
\usepackage{ae,aecompl}
\usepackage{graphicx}
\usepackage[]{natbib}
\usepackage{amsmath}
\usepackage{float}
\usepackage{deluxetable}
\bibpunct{(}{)}{,}{a}{}{,}

\def\ffc{MACS~J1149.5$+$2223}
\def\ffd{MACS~J0717$+$3745}
\def\ffe{Abell 370}
\def\fff{RXC~J2248.7$-$4431}
\def\ffas{A2744}
\def\ffbs{MACS0416}
\def\ffcs{MACS1149}
\def\ffds{MACS0717}
\def\ffes{A370}
\def\fffs{RXJ2248}
\def\HST{{\it HST}}

\def\Spitzer{\it Spitzer}

\def\abs#1{\left\vert #1 \right\vert}

\def\chone{3.6\:\mu\rm{m}}
\def\chtwo{4.5\:\mu\rm{m}}

\newcommand{\hst}{\it HST}
\newcommand{\surfsup}{SURFSUP}         

\title[Hubble Frontier Field Photometric Catalogues of {\ffe} and
{\fff}]{Hubble Frontier Field Photometric Catalogues of {\ffe} and {\fff}: Multiwavelength
  photometry, photometric redshifts, and stellar properties}

\author[Brada\v{c} et al.]{Maru\v{s}a Brada\v{c},$^{1}$\thanks{E-mail: marusa@physics.ucdavis.edu}
Kuang-Han Huang,$^{1}$
Adriano Fontana,$^{2}$
Marco Castellano,$^{2}$
\newauthor 
Emiliano Merlin,$^{2}$
Ricardo Amor\'{i}n,$^{3,4}$
Austin Hoag,$^{1,5}$
Victoria Strait,$^{1}$
\newauthor
Paola Santini,$^{2}$
Russell~E.~Ryan~Jr.,$^{6}$
Stefano Casertano,$^{6}$
Brian C. Lemaux,$^{1}$
\newauthor 
Lori M. Lubin,$^{1}$
Kasper B. Schmidt,$^{7}$ 
Tim Schrabback,$^{8}$ 
Tommaso Treu,$^{5}$
\newauthor Anja von der Linden,$^{9}$
Charlotte A. Mason,$^{5,10}$
and Xin Wang.$^{5}$
\\
$^{1}$Department of Physics, University of California, Davis, CA 95616, USA\\
$^{2}$INAF - Osservatorio Astronomico di Roma Via Frascati 33 - 00040
Monte Porzio Catone, 00040 Rome, Italy\\
$^{3}$ Instituto de Investigaci\'on Multidisciplinar en Ciencia y Tecnolog\'ia, Universidad de La Serena, Ra\'ul Bitr\'an 1305, La Serena, Chile\\
$^{4}$ Departamento de F\'isica y Astronom\'ia, Universidad de La Serena, Norte, Av. Juan Cisternas 1200, La Serena, Chile \\
$^{5}$Department of Physics and Astronomy, UCLA, Los Angeles, CA, 90095-1547, USA\\
$^{6}$ Space Telescope Science Institute, 3700 San Martin Drive, Baltimore, MD 21218, USA\\
$^{7}$Leibniz-Institut f\"{u}r Astrophysik Potsdam (AIP), An der
Sternwarte 16, 14482 Potsdam, Germany\\
$^{8}$Argelander-Institut f\"{u}r Astronomie
Auf dem H\"{u}gel 71
D-53121 Bonn, Germany\\
$^{9}$Stony Brook University, Department of Physics and Astronomy, Stony
Brook, NY 11794, USA\\
$^{10}$Center for Astrophysics, Harvard \& Smithsonian, 60 Garden St,
Cambridge, MA, 02138, USA; Hubble Fellow}

\begin{document}
\label{firstpage}
\pagerange{\pageref{firstpage}--\pageref{lastpage}}
\maketitle
\begin{abstract}  
  This paper presents multiwavelength photometric catalogues of the last two Hubble
  Frontier Fields (HFF), the massive galaxy clusters {\ffe} and
  {\fff}. The photometry ranges from imaging performed on the
  \emph{Hubble} Space Telescope (HST) to ground based Very Large
  Telescope (VLT) and
  \emph{Spitzer}/IRAC, in collaboration with the ASTRODEEP team, and
  using the ASTRODEEP pipeline. While the main purpose of this paper is to release the
  catalogues, we also perform, as a proof of concept, a
  brief analysis of $z>6$ objects selected using drop-out method, as
  well as spectroscopically confirmed sources and multiple images in
  both clusters. While dropout methods yield a sample of high-z
  galaxies, the addition of longer wavelength data reveals that as
  expected the samples have substantial contamination at
  the $\sim$30-45\% level by dusty galaxies at lower
  redshifts. Furthermore, we show that spectroscopic redshifts are still required
  to unambiguously determine redshifts of multiply imaged
  systems. Finally, the now publicly available ASTRODEEP catalogues were combined for all HFFs and used to explore stellar properties
  of a large sample of 20,000 galaxies across a large photometric
  redshift range. The powerful magnification provided by the HFF
  clusters allows for an
  exploration of the  properties of galaxies with intrinsic stellar masses as low as
  $M_{\ast} \gtrsim 10^7M_{\odot}$ and intrinsic star formation rates
  SFRs$\sim 0.1 \mbox{--} 1 M_{\odot}/\mbox{yr}$ at $z>6$.
\end{abstract}



\section{Introduction}
\label{sec:intro}

The Hubble Frontier Field campaign is a multi-cycle observing campaign
using Director's Discretionary Time with Hubble Space Telescope (HST) and Spitzer Space
Telescope to study the faintest galaxies. It is particularly suited to
observe typical (i.e. sub-$L^*$, where $L^*$ is the characteristic
luminosity) galaxies at high redshifts. To achieve this, the Frontier
Fields combine the power of HST with the gravitational telescopes:
six high-magnification clusters of galaxies.  Abell 2744,
MACSJ0416.1-2403, MACSJ0717.5+3745, MACSJ1149.5+2223, Abell 370, and  RXCJ2248.7-4431
(also known as Abell S1063) have been targeted in the optical by the
HST Advanced Camera for Surveys (ACS) and the infra red Wide Field Camera 3 (WFC3/IR)  with coordinated parallel fields for
over 840 HST orbits. This data is complemented with the data from
previous surveys (e.g, Cluster Lensing And Supernova survey with
Hubble; CLASH; \citealp{postman12}). The Spitzer Space
Telescope also dedicated Director's Discretionary Time to obtain IRAC
${\chone}$ and ${\chtwo}$ imaging to achieve the total exposure of
50hr/band/cluster.  The {\Spitzer} data for some of the clusters are complemented as well by
data from previous surveys (mainly Spitzer UltRa Faint Survey Program
{\surfsup}, \citealp{surfsup}). Deep
Ks images from VLT High Acuity Wide field K-band Imager (HAWK-I) are also included \citep{brammer16}.

\begin{figure*}
  \begin{minipage}{1.0\textwidth}%
\begin{minipage}{0.5\textwidth}%
  \includegraphics[width=1\textwidth]{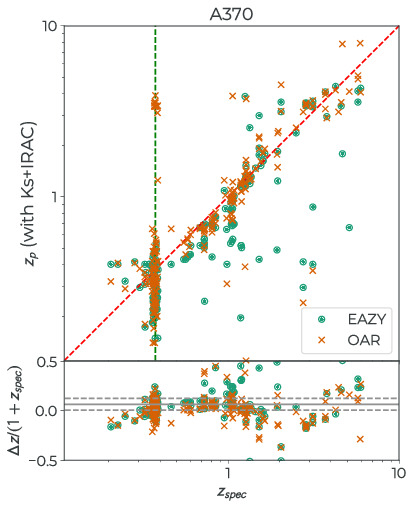}
\end{minipage}%
\begin{minipage}{0.5\textwidth}
  \includegraphics[width=1\textwidth]{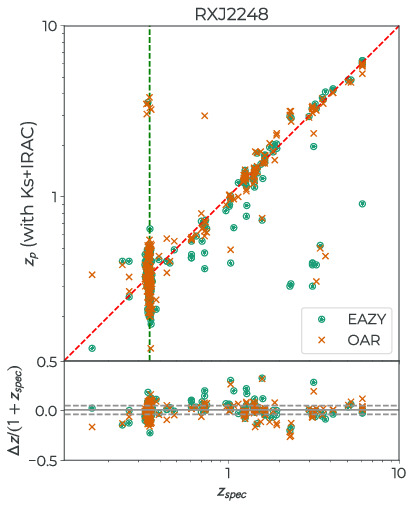}
\end{minipage}%
\end{minipage}%
\caption{Comparison of the {\ffes} (left) and {\fffs} (right) spectroscopic redshifts compiled from the
    literature by \citet{shipley18} vs. photometric redshifts derived in this work using full
     photometry (HST+VLT/HAWK-I+Spitzer). Circles and crosses present
   the two methods used for photometric redshifts, EAZY and OAR, respectively. The
   bottom panels show the residuals, 
   solid and dashed line are the median and standard deviation
   respectively (see also Table~\ref{tab:stat}). The increased ICL
   component of {\ffes} likely leads to the lower accuracy/precision
   of the photometric redshifts for both methods.}
    \label{fig:zspec}
  \end{figure*}
  
\begin{figure}
\centerline{\includegraphics[width=.5\textwidth]{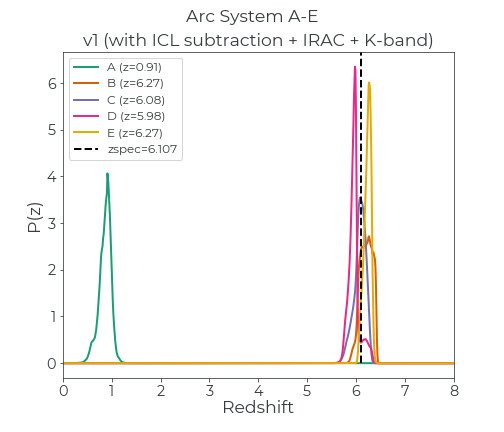}}
  \caption{Comparison of our estimates of the photometric redshifts distribution for a quintuply
    imaged system at $z_{\rm spec}= 6.107$ behind {\fff} \citep{karman15,schmidt17}. Plotted are photometric redshift probability distributions
     using ICL subtraction and full
     photometry (HST+VLT/HAWK-I+Spitzer) for each of the images A-E. Their spectroscopic
     redshifts are given by a dashed line, and best photometric
     redshifts are indicated in the legend. Image A is located
very close to the core of the cluster and its photometry is less precise \citep{schmidt17}.}
    \label{fig:arcsRXC}
 \end{figure}

\begin{figure*}
  \begin{minipage}{0.5\textwidth}
    \includegraphics[width=1.0\textwidth]{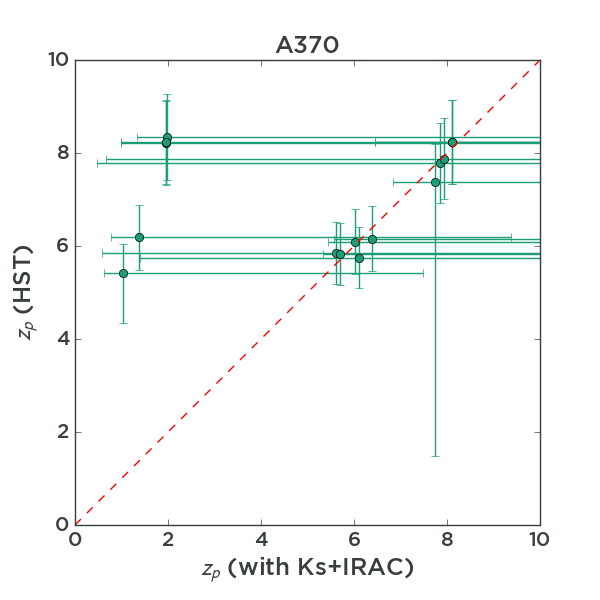}
    \end{minipage}%
    \begin{minipage}{0.5\textwidth}
      \includegraphics[width=1.0\textwidth]{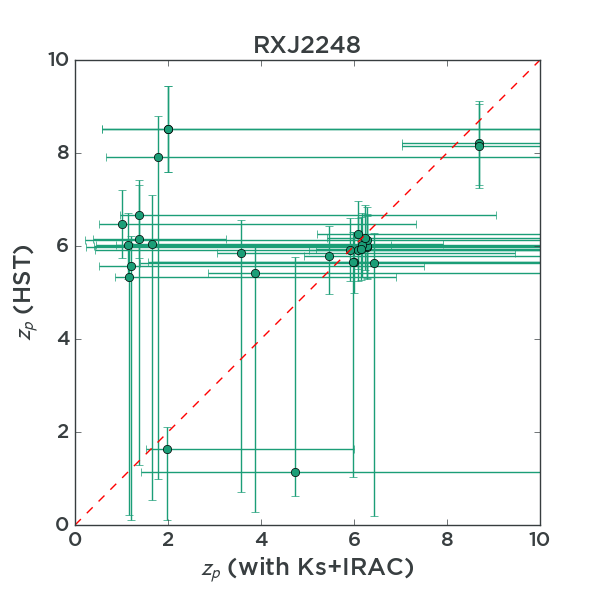}
\end{minipage}%
  \caption{Comparison of the redshifts for the high redshift dropouts behind
     {\ffes} (left) and   {\fffs} (right) from
     \citet{ishigaki17}. Plotted are photometric redshifts and errors from
     \citet{ishigaki17} using only HST data vs. EAZY redshifts and
     95\% confidence limits derived in this work using full
     photometry (HST+VLT/HAWK-I+Spitzer).}
    \label{fig:zphot}
 \end{figure*}

\begin{table} 
 \caption{Photomeric redshift accuracy. Listed are biweight location of the 
$(z_{\rm spec}-z_{\rm phot})/(1+z_{\rm spec})$, median absolute
deviation $\sigma_{\frac{\Delta z}{1+z_{\rm spec}}}$ and
number of outliers defined as $\abs{(z_{\rm spec}-z_{\rm
    phot})/(1+z_{\rm spec})}>0.15$. }\label{tab:stat}
\begin{tabular}{lcccc}
\hline   
Cluster & Sample &
$\left <\frac{\Delta z}{1+z_{\rm spec}} \right >$ &
$\sigma_{\frac{\Delta z}{1+z_{\rm spec}}}$ & Outliers\\
\hline \hline
{\ffes} EAZY & 202 & 0.06 & 0.05 & $21\%$\\
{\ffes} OAR & & 0.02 & 0.07 & $24\%$\\
\hline
{\fffs} EAZY & 210 & 0.0006 & 0.04 & $9\%$\\
{\fffs} OAR & & 0.005 & 0.04 & $9\%$\\
\hline
\end{tabular}
\end{table}

The main high level science products that make rich data sets such as those in the HFFs
 even more useful for the community are photometric catalogues that
combine all the available imaging in a consistent manner. Photometric catalogues for the first four clusters have been
published and provided to the community \citep{merlin16,castellano16,
  dicriscienzo17}. In collaboration with the ASTRODEEP team, we
provide equivalent catalogues for the last two HFF clusters {\ffe} (hereafter
{\ffes}) and {\fff} (hereafter {\fffs}) using almost identical methods
to those employed for the first four HFF clusters
\citep{merlin16,castellano16, dicriscienzo17, santini17}. Though catalogues have
also been published by \citet{shipley18} for all six HFF clusters, the
catalogues presented here use a different methodology for measuring
photometry, photometric redshifts, and stellar properties; therefore they provide independent and
complementary measurements. We use the spectroscopic catalogues assembled by
\citet{shipley18}, as well as perform some high-level comparisons
throughout the paper.

In this paper we describe the new catalogues and investigate the utility
of the longer wavelength data by investigating the high-redshift
dropout candidates. In addition, we also perform comparison of
photometric redshifts with known
spectroscopic redshifts,  including for multiply imaged sources.  Finally, we combine data for all six HFF clusters and explore stellar
  properties of a large sample of 20,000 galaxies. The paper is structured as follows. In Section~\ref{sec:data} we
present the data used to generate the catalogues and in
Section~\ref{sec:dataanalysis} we describe the steps taken to generate
these catalogues and their public release. In
Section~\ref{sec:results} we present the main science results that include
redshift comparisons and measurements of stellar properties.  We summarize in
Section~\ref{sec:conclusions} and give the location of publicly
released catalogues in Appendix~\ref{sec:app}.

Throughout the paper we assume a
$\Lambda$CDM concordance cosmology with $\Omega_{\rm m}=0.3$,
$\Omega_{\Lambda}=0.7$, and Hubble constant
$H_0=70{\rm\ kms^{-1}\:\mbox{Mpc}^{-1}}$
\citep{komatsu11, riess11}.
Coordinates are given for the epoch J2000.0, and magnitudes are in the
AB system.

\section{Data}
\label{sec:data} {\ffes} and {\fffs} are the final two clusters from the
HFF campaign. They were imaged with 140 orbits each in three optical
(ACS; F435W, F606W, F814W) and four near infra-red (WFC3; F105W,
F125W, F140W, F160W) bands in one pointing. We use the {\hst} data
available  here\footnote{\url{http://www.stsci.edu/hst/campaigns/frontier-fields/FF-Data}},
in
particular we use version 1.0 epochs 1 and 2 in both cases. In
order to combine it with {\Spitzer} data we use images drizzled to
$0.06\arcsec/{\rm pixel}$ scale. We use the {\Spitzer} data
available here
\footnote{\url{http://irsa.ipac.caltech.edu/data/SPITZER/Frontier/}}
as well as tools that were developed for our {\surfsup} program
(\citealp{surfsup,surfsup2,huang15a};  HFF postdates {\surfsup} and
only {\ffc} and {\ffd} data is in common). Finally we also use HAWK-I data from
the VLT/ESO program 092.A-0472(A) (PI Brammer,
\citealp{brammer16}) and spectroscopic data from Keck/LRIS,
VLT/MUSE, VLT/FORS2, Magellan/LDSS3, Keck/Deimos,
and HST/GRISM (GLASS program) collated by \citet{shipley18} using various literature
sources (\citealp[][in
prep.]{brammer18}, \citealp{lagattuta17,treu15, karman17, diego16, richard14}). 

\begin{figure*}
\begin{minipage}{0.5\textwidth}
\includegraphics[width=0.95\textwidth]{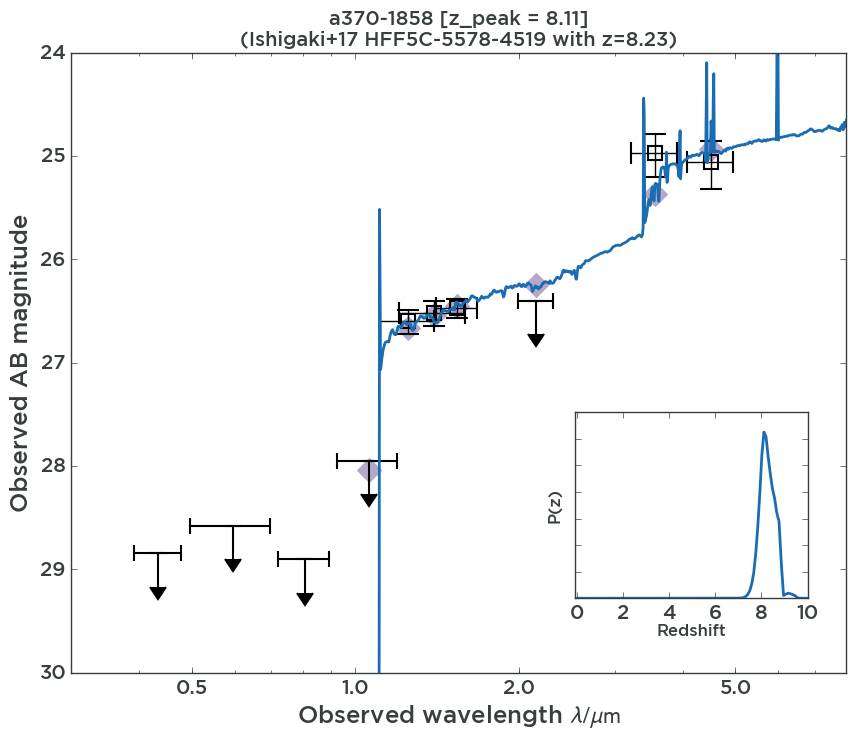} 
  \end{minipage}%
  \begin{minipage}{0.5\textwidth}
\includegraphics[width=0.95\textwidth]{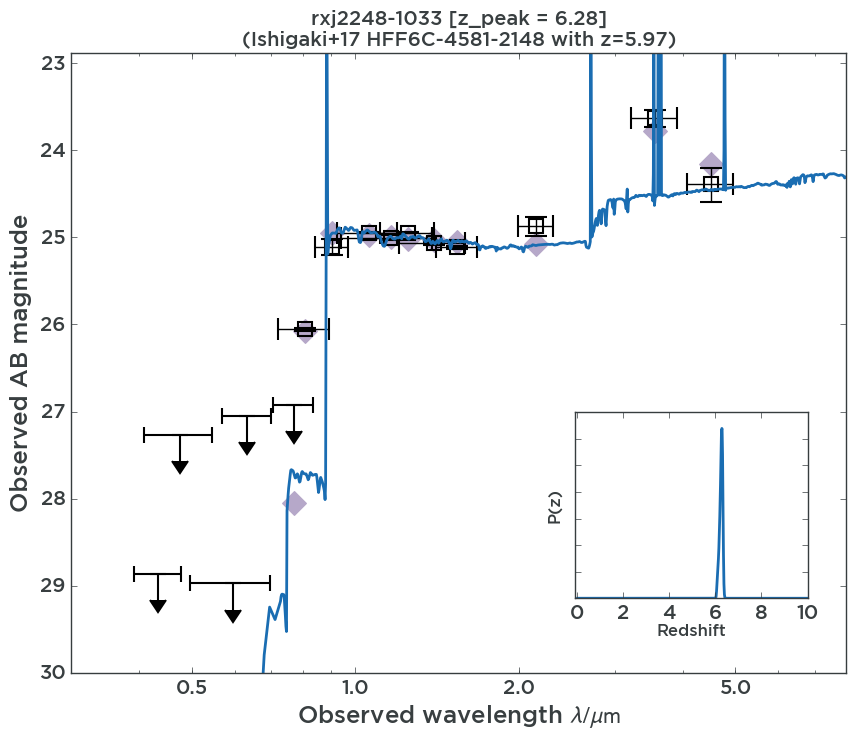} 
    \end{minipage}%
   \caption{Examples of spectral energy distribution (observed AB
     magnitude vs wavelength) and redshift probability
     distribution (insets) plots for two objects behind
     {\ffes} (left) and  {\fffs} (right). Points with error bars show multiband
     photometric data from this work, line shows the best fit
     template as fit by EAZY and diamonds represent the expected
     magnitudes based on this template. Both objects are selected to
     be high redshift galaxies based on their photometry. In addition,
     both show IRAC detections indicating either strong nebular
     emission lines or old stellar populations based on their Balmer breaks.}
    \label{fig:sed}
 \end{figure*}

\section{Data Analysis and Catalogues}
\label{sec:dataanalysis}
Our data analysis closely follow the procedures outlined in \citet{merlin16,castellano16, dicriscienzo17, huang16}. For completeness, we briefly outline the procedure below. 

To improve detection of faint sources in the cluster, we start by
modeling and subtracting diffuse intra-cluster light (ICL) in the HST
F160W images using the procedure outlined in \citet{merlin16}. This is
to remove the spatially varying background in the cluster field that
complicates photometry (especially for faint, high-redshift sources
that we are targeting). We first mask out bright pixels above 8 times
the estimated sky level, and then we use GALFIT \citep{peng11} to
model the ICL with one component using Ferrer profiles
\citep{giallongo14}. The initial guesses for the centroid, central
surface brightness, and truncation radius are the cluster center
(brightest cluster galaxy), $22 \mbox{mag/arcsec}^2$, and $30\arcsec$, respectively. The purpose of fitting ICL with Ferrer profile is not to carefully characterize ICL (as in, e.g., \citealp{morishita17b}), but rather to obtain images with more uniform background for photometry. 

After we obtain an initial estimate of the ICL component, we fix the
ICL parameters and use GALAPAGOS \citep{barden12} and GALFIT to model
bright cluster galaxies. This step involves a first run of SExtractor  \citep{sextractor}
to obtain initial guesses for GALFIT for each bright cluster member as
well as adding secondary GALFIT components to each cluster member to
better model their light profiles (especially around the cores). This
step is particularly important for A370; because of its low redshift
compared with other clusters in the HFF sample, its cluster members
occupy a larger fraction of the field of view and make detecting
background high-redshift sources more challenging. After satisfactory
models of bright cluster members are obtained, we refine the ICL
component by relaxing its centroid position, central surface
brightness, and truncation radius. Although
subtracting ICL and bright cluster members does not improve the
signal-to-noise ratios of faint sources, it makes detecting them using
SExtractor easier by reducing gradients in local background. It is
also a lot easier to visually assess the detection of faint sources
once ICL and bright cluster members are removed.

After the above process is finished for F160W (our detection image),
we repeat the same process for all other HST filters, using the
best-fit parameters from the next redder filter as initial
guesses. Modeling of ICL and bright cluster members are done
separately on IRAC and $K_s$ bands because of their lower-resolutions,
which requires a different tool (T-PHOT) as explained below.

We extract photometry on the HST images using
SExtractor. For the final detection catalogues we
use F160W processed
images and use SExtractor with a HOT+COLD approach
\citep{galametz13}. This procedure adopts two different sets
of the SExtractor parameters to detect objects at different spatial
scale, COLD for bright extended objects and HOT for faint galaxies. 
We also match the point-spread functions (PSFs) among all HST filters to
get consistent colour. To this aim, we identify isolated point sources in
each cluster field, and we use the {\tt psfmatch} task in {\tt IRAF}
to match all HST images to have the same PSF as the F160W band.

To determine the {\Spitzer}-{\hst} and VLT/HAWKI-{\hst} colors we use
the template fitting software T-PHOT \citep{merlin15,merlin16b}. This is
necessary, as unlike the PSF between different {\hst} images, the PSF
of especially {\Spitzer}/IRAC is much larger ($\sim 2\arcsec$)
compared to {\hst} ($\sim 0.1\arcsec$). To prepare the HST images for
T-PHOT, we use the public $0.06\arcsec/{\rm pixel}$ scale images. We
also edit the astrometric image header values (CRVALs and CRPIXs, see \citealp{merlin15}) to
conform to T-PHOT's astrometric requirements and make
    sure that HST and {\Spitzer} images are aligned to well within
    $0.1\arcsec$. 

Finally we use T-PHOT to
    measure the fluxes in the low-resolution image (in our case the
    IRAC and VLT/HAWK-I Ks images) for all the sources detected in the high-resolution
    image (in our case with the F160W {\hst} images). T-PHOT does so by
    constructing a template for each source; it convolves the cutout
    of each source in the F160W image with a PSF-transformation kernel
    that matches the F160W resolution to the IRAC resolution. T-PHOT
    solves the set of linear equations to find the
    combination of coefficients for each template that most closely
    reproduces the pixel values in the IRAC image. 
    Finally, all fluxes
    are collated in our final combined photometric catalogues (see Appendix~\ref{sec:app}).

We determine photometric redshifts using two different photometric redshift
codes 1) EAZY (Eazy and Accurate Zphot from Yale, \citealp{brammer08})  and 2) OAR
(Osservatorio Astronomico di Roma, \citealp{fontana00}) code. We use EAZY with the \citet[][hereafter BC03]{bc03}
templates. For this procedure we set a minimum allowed photometric
uncertainty corresponding to 0.05 mags for the HST and
HAWK-I bands and 0.1 mags for the IRAC bands: errors smaller
than these values are replaced by the minimum allowed uncertainty to
account for the zero-point uncertainties. We use the redshifts that
correspond to the maximum likelihood probability in our final
solution. We
account for dust attenuation internal to the galaxy following the
prescription by \citet{calzetti00}. The templates also include strong
nebular emission lines, whose fluxes are determined by the
Lyman continuum flux of BC03 models and nebular line ratios
from \citet{anders03}.

The OAR photometric redshifts are
obtained with the {\tt zphot.exe} code \citep{fontana00} following the 
procedure described by  \citet{grazian06} (see also \citealp{dahlen13, santini15}). Best-fit
photo-zs are obtained through a $\chi^2$ minimization  using SED templates from PEGASE 2.0
\citep{pegase}. For this procedure we also set minimum photometry
errors as described above. Throughout this work we use EAZY
photometric redshifts, except when comparing with the spectroscopic
sample where we use both (Sect.~\ref{sec:zspec}). Note that in neither
case do we assume a prior to account for the existence of
each cluster, in doing so photometric redshifts at the cluster
redshift would improve \citep{morishita17}.

\begin{figure*}
\includegraphics[width=.7\textwidth]{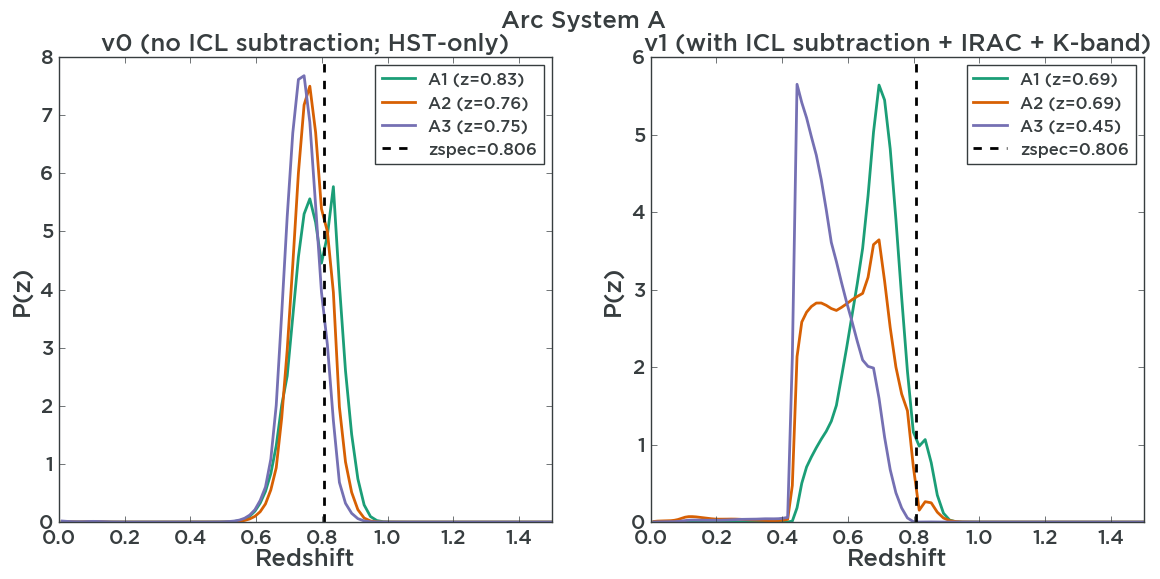}\\
\includegraphics[width=.7\textwidth]{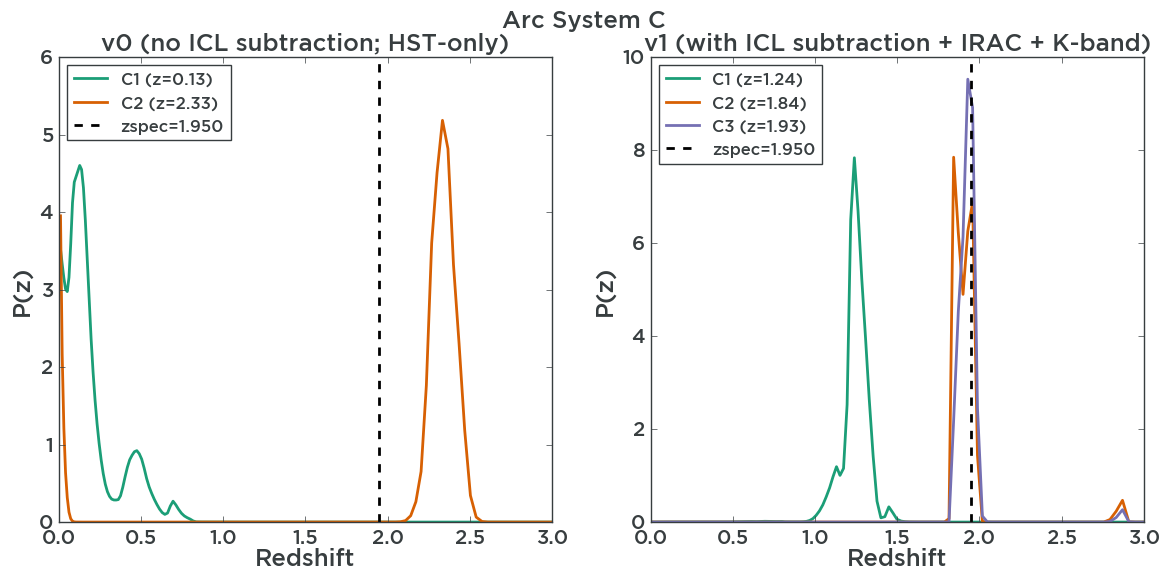}\\
\includegraphics[width=.7\textwidth]{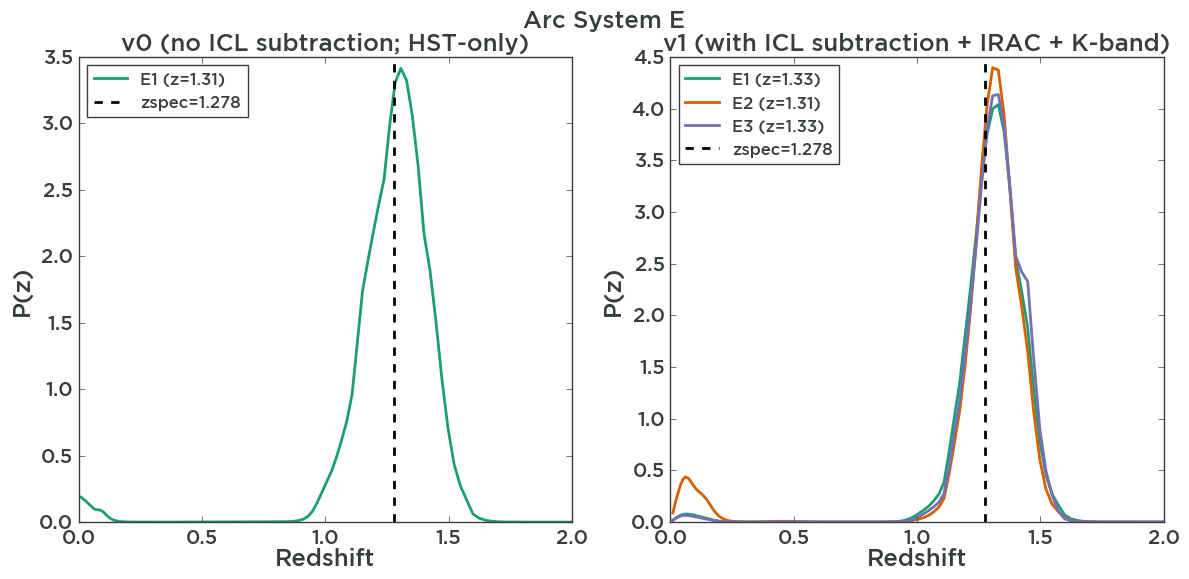}\\
  \caption{Comparison of the photometric redshifts for the multiply imaged systems behind
     {\ffe}. Plotted are photometric redshift probability distributions
     using only HST photometry without ICL subtraction and redshifts
     derived in this work using ICL subtraction and full
     photometry (HST+VLT/HAWK-I+Spitzer). Three systems with measured 
     spectroscopic redshifts A (top, system 1 in \citealp{strait18}), C
     (middle, system 3) and E (bottom, system 5) are plotted and their spectroscopic
     redshifts are given by a dashed line. While ICL subtraction
     allows us to detect fainter images of the system (case C and E),
     even with the multiband photometry the photometric redshift solution
     can still lead to incorrect estimates of the redshift of the
     system (as is the case for system A, which is
     located close to the cluster core and its photometry is less
     reliable).}
    \label{fig:arcsA370}
 \end{figure*}

Galaxy physical properties are computed as described by \citet{castellano16} fitting
BC03 templates with the
{\tt zphot.exe} code at the previously determined spectroscopic
redshift where available or photometric
redshift from the EAZY code ($z_{\rm best}$). Only sources with
reliable redshifts $z_{\rm best}\ge 0$ that have
reliable photometry (no artifacts and coverage in most of the
bands) are used. Using OAR photometric redshift does not significantly
change the results. For this cursory analysis, to allow for the
broadest possible comparison, we adopt a suite of SFHs most commonly
employed during the SED process for deep extragalactic surveys. In the BC03 fit we assume exponentially declining
star-formation histories (SFHs) with e-folding time $0.1 \leq \tau \leq
15.0\mbox{Gyr}$. Note, however, that stellar masses are only mildly
sensitive to the choice of the SFH \citep{santini15}, and this choice does not
significantly affect our results. We assume a \citet{salpeter55} initial mass function and we allow both
\citet{calzetti00} and Small Magellanic Cloud \citep{prevot84} extinction laws. Absorption by the intergalactic medium
(IGM) is modeled following \citet{fan06}. We fit all the
sources with stellar emission templates including
the contribution from nebular continuum and line emission
following \citet{schaerer09} under the assumption of
an escape fraction of ionizing photons $f_{\rm esc} = 0.0$ (see also
\citealp{castellano14} for details). SFRs were estimated from UV
rest-frame photometry using approach outlined in
\citep{castellano12}. UV slope $\beta$ was used to obtain the the
dust-corrected UV magnitude, which is then used to obtain an SFR
estimate with the \citealp{kennicut12} factor.
We also release catalogues of these
properties as
described in the Appendix~\ref{sec:app}.

\section{Results}
\label{sec:results}
\subsection{Comparison with Spectroscopic samples}\label{sec:zspec}

After the photometric catalogues were finalized, we compared
spectrocopic redshifts to photometric redshifts as computed using
methodology described above. We did not use
spectroscopic redshifts to adjust the imaging zero-points. In
addition, the photometry was not optimized for large galaxies
(i.e. cluster members, see e.g. \citealp{tortorelli18}), as our primary goal was to study high-redshift
galaxies.   Spectroscopic redshifts were
recently collected by \citet{shipley18} using various literature
sources. For {\ffes} the catalogues are \citet[][in
prep.]{brammer18}, \citet{lagattuta17,treu15, richard14} and for  {\fffs} they used \citet[][in
prep.]{brammer18}, \citet{ karman17, diego16, treu15, richard14}. The
comparison is given in
Figure~\ref{fig:zspec}. Overall, the
photometric redshift performance is very similar to the performance
reported by 
\citet{shipley18,castellano16, dicriscienzo17}. The results for the
biweight location (a robust statistic for determining the central location of a distribution) of the 
$(z_{\rm spec}-z_{\rm phot})/(1+z_{\rm spec})$, median absolute deviation and
number of outliers defined as $\abs{(z_{\rm spec}-z_{\rm
    phot})/(1+z_{\rm spec})}>0.15$ are listed in
Table~\ref{tab:stat}. The fraction of catastrophic outliers is higher for {\ffes}, likely due to larger
ICL contamination. From now on, unless specified otherwise, we will
use EAZY photometric redshifts.

\subsection{High Redshift Galaxies}\label{sec:dropouts}
One of the main goals of the HFF program was to detect high redshift,
highly magnified galaxies. We briefly perform an analysis here to
investigate galaxies with secure spectral redshifts at $z>6$. For the $z>6$
population very few spectroscopic redshifts exist. For the two clusters
studied in this work we have a total of 2 galaxies with spectroscopic
redshifts  at
$z>6$. These are $z=6.5$ object by
\citet{hu02} behind {\ffes} and a quintuply imaged system at $z=6.107$ behind
{\fffs}  \citep{karman15,schmidt17}. For {\fffs}, four images are correctly identified at
$z=6.107$ within $\pm 0.2$, while one fails catastrophically and is put
at $z\sim 1$ (Fig.~\ref{fig:arcsRXC}). The image that fails is located
very close to the core of the cluster and its photometry is likely
affected. In 
\citet{shipley18} one of the objects also fails (a different one) and is put at $z\sim 4$.

We also looked into the sample from \citet{ishigaki17},
where high redshift galaxies were selected based on the dropout
technique \citep{steidel96}, and their photometric redshifts were
determined subsequently using only HST data. The dropout technique is based on the
photometric detection/non-detection of objects near the Lyman
break. As such it does not use rest-frame optical information. The results are shown in
Fig.~\ref{fig:zphot}. The addition of rest-frame optical data
(HST+VLT/HAWK-I+Spitzer) is essential, as it can often identify lower
redshift dusty objects for which the break could mimic the
Lyman$\alpha$/Lyman-limit break.
Hence, we see a non-trivial fraction ($\sim$30-45\%) of objects that scatter to lower
redshift when such data is added. {\Spitzer} data is especially powerful
in this case, as it targets high equivalent width nebular emission
lines and/or can detect ``old'' stellar populations based on the 
     4000${\mbox{\AA}}$ break (see Fig.~\ref{fig:sed} for examples of
     SED fitting). This not only improves accuracy of redshift determination, but also allows us to better study stellar properties
     at highest redshifts.

\subsection{Multiple Imaged Systems}
\label{sec:arcs}

Another common application of photometric redshifts is to determine
redshifts of the multiply imaged systems to be used for strong
gravitational lensing and accurate determinations of projected
mass distribution and magnification of clusters. This is important
for high redshift studies, as stellar masses and SFRs need to be
corrected for magnification to obtain intrinsic values (see Sect.~\ref{sec:sprop}). Erroneous
redshifts can significantly bias results (e.g.,
\citealp{treu16,grillo16,remolinagonzalez18}). We look into how well
photometric redshifts fulfill this task for the set of multiple images
with spectroscopic redshifts.  These are some of the more difficult
objects on which to perform accurate photometry on. They are often
distorted, hence traditional photometry approaches can fail. Our
results are shown in Figs.~\ref{fig:arcsRXC}, \ref{fig:arcsA370}. While ICL subtraction
allows us to detect fainter images of the system (case C and E in Fig.~\ref{fig:arcsA370}), even with the full multiband photometry the photometric
redshift solution can be biased. An important quantity to consider is
the angular diameter distance ratio between the source and the lens
(deflector) and the observer and the source $D_{\rm ds}/D_{\rm
  s}$. For a typical lens at redshift $z_{\rm d}=0.5$, this
ratio changes by $10\mbox{-}50\%$ for source redshifts of
$z_{\rm s} = 1.3\mbox{--}0.7$ assuming source redshift error of
$\Delta z/(1+z)=0.1$. The error on $D_{\rm ds}/D_{\rm
  s}$ (if present in the same
direction for all multiple images) will then directly translate to the
error in normalization of the mass distribution. Therefore, whenever
performing lens modeling it is best to obtain a large spectroscopic
sample.

\begin{figure}
 \centerline{\includegraphics[width=0.5\textwidth]{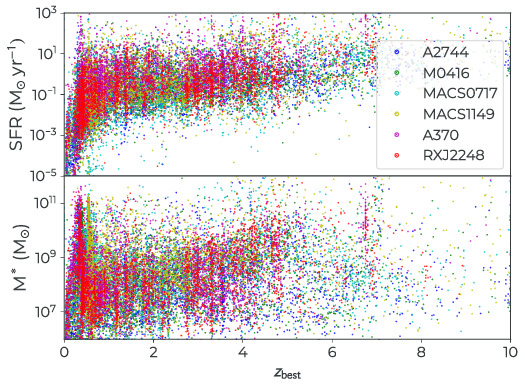}}
\centerline{\includegraphics[width=0.5\textwidth]{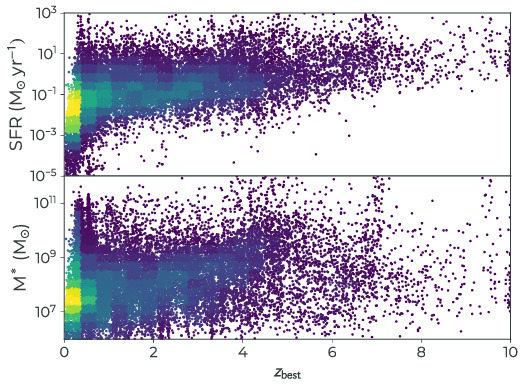}}
\caption{Star Formation Rate (SFR) and stellar mass ($M_{\ast}$)
 sepparated by cluster ({\bf top}) and as a hybrid density plot ({\bf
   bottom}) as a function of best redshift $z_{\rm best}$. The plots
 include all six HFF clusters (main
  pointing, {\ffas}, {\ffbs} from \citealp{castellano16}, {\ffcs},
  {\ffds} from \citealp{dicriscienzo17}, and {\ffes}, {\fffs} from
  this work). Both SFR and $M_{\ast}$ have been corrected for magnification
  using median magnifications from all submitted lens models as
  described by \citet{castellano16}. 
  Gravitational lensing allows galaxies to be detected at
 stellar masses as low as $10^7M_{\odot}$ and intrinsic $\mbox{SFRs}\sim 0.1\mbox{--}1 M_{\odot}/\mbox{yr}$ even at the highest redshifts.}
    \label{fig:SFRmass}
 \end{figure}

\subsection{Stellar Properties}
\label{sec:sprop}

\begin{figure}
\includegraphics[width=0.5\textwidth]{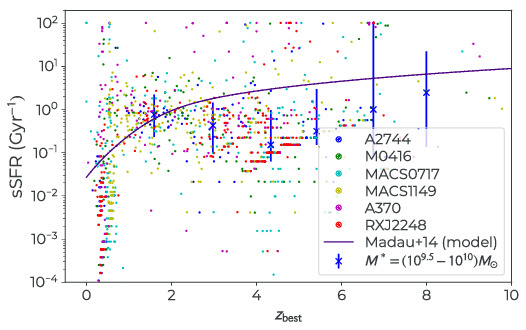}
  \caption{sSFR as a function of redshift $z_{\rm best}$ for all 6 HFF clusters (main
  pointing, {\ffas}, {\ffbs} from \citealp{castellano16}, {\ffcs},
  {\ffds} from \citealp{dicriscienzo17}, and {\ffes}, {\fffs} from
  this work). Also plotted is the theoretical model from
  \citet{madau14} and binned data points with errorbars. Only galaxies
  with $M_{\ast}=10^{9.5}-10^{10}M_{\odot}$ are plotted and
considered in the binning of the data as is often done in the
literature. The maximum values of
$\mbox{sSFR}=10^2\mbox{Gyr}^{-1}$ are indicative of the youngest stellar
population models we use ($10\mbox{Myr}$). Note that sSFR is independent of magniffication.     \label{fig:sSFR}}
 \end{figure}

Stellar properties of galaxies in all six catalogues
(\citealp{castellano16, dicriscienzo17} and this work) are presented
in Figs.~\ref{fig:SFRmass},\ref{fig:sSFR}.  Each cluster's main
pointing contains $3000-4000$ galaxies with measured properties for a
total of $20,000$ objects.  In Fig.~\ref{fig:SFRmass} we show Star
Formation Rate (SFR) and stellar mass ($M_{\ast}$) as a function of
redshift. Both quantities have been corrected for magnification using
median magnifications from version 4 lens models as described by
\citet{castellano16} (see models and version description here
\footnote{\label{fn:lm}\url{https://archive.stsci.edu/prepds/frontier/lensmodels/}}).  It is
very encouraging to see that we can target galaxies down to stellar
mass of $10^7 M_{\odot}$ even at the highest redshift (this is similar
to the mass of Fornax dwarf spheroidal, \citealp{kirby13}). At
intermediate redshifts some stellar mass might be coming from
relatively evolved ($\sim 500\mbox{Myr}\mbox{--}1\mbox{Gyr}$)
populations due to the lack of rest frame near-IR data and in that
case, these low masses should be considered as lower limits. However,
at the highest redshifts any contribution from dusty populations is
likely to be sub-dominant. Similarly we can detect galaxies down to
intrinsic $\mbox{SFRs}\sim 0.1\mbox{--}1 M_{\odot}/\mbox{yr}$ at $z>6$.

Fig.~\ref{fig:sSFR} shows a plot of a specific star formation rate (sSFR) as a function of
$z_{\rm best}$.  These results  are independent of lens
magniffication. However, it is the magnification which enables us to obtain a more
complete sample down to lower stellar masses. The maximum values of
$\mbox{sSFR}=10^2 \mbox{Gyr}^{-1}$ are indicative of the youngest stellar
population models we use ($10\mbox{Myr}$). We only plot galaxies with $M_{\ast}=10^{9.5}\mbox{--}10^{10}M_{\odot}$ as is often done in the
literature (e.g., \citealp{santini17}) and $68\%$ confidence limits with median value in each bin. The results are consistent with
e.g. \citet{tasca15b} at $2 < z < 5$; though \citet{tasca15b} sample includes only galaxies with spectroscopic
redshifts and thus has a cleaner sample.

Qualitatively at high redshifts our results are affected by
incompleteness in stellar mass (we are less likely to detect low
stellar mass objects).  Since F160W traces rest-frame UV light, only
high SFR objects will enter our sample. In order to estimate the
incompleteness in SFR we would need a complete sample of galaxy
colours at high redshifts to estimate the full range of SFRs; such a
sample is not available. In addition, selecting galaxies based on rest
frame optical data is not possible at present due to the relatively
shallow depth and large PSF of the {\Spitzer} data. The measurement
errors also increase for high redshift and faint sources, which could
lead to the Eddington bias. As shown by \citet{santini17}, correcting
for the Eddington bias would increase sSFR at $z>3$.  Finally, as with
any sSFR measurement the systematic uncertainty of measuring SFR
(e.g., lack of direct tracers such as dust corrected $H\alpha$,
uncertainties due to unknown IMF) and $M_{*}$ (e.g., uncertainties
due to unknown IMF) using photometry remain. The detailed explorations
of sSFR at highest redshifts will thus have to await the launch of
{\it JWST}.

\section{Conclusions} \label{sec:conclusions}
In this paper we present and publicly release photometric catalogues
of two HFF clusters, {\ffe} and {\fff}. The catalogues
include {\HST}, HAWK-I/Ks band and {\Spitzer} data. We measure photometric
redshifts for all sources and compare them to spectroscopic
data  from the literature. Comparison shows a reasonable agreement with $\sigma_{\Delta
  z/(1+z_{\rm spec})} \sim 0.05$ and an outlier fraction of
$10-20\%$. The fraction is higher for {\ffes}, likely due to larger
ICL contamination. We have also explored the accuracy of photometric
redshifts for strongly lensed systems and conclude that their errors
can cause a significant bias in lens modeling.

Finally, we explore the stellar properties of galaxies using samples from all
6 HFF clusters, containing 20,000 galaxies. The magnification from a
foreground cluster 
allows for the detection of objects with stellar mass $M_{\ast} \gtrsim
10^7M_{\odot}$ and intrinsic SFRs$\sim 0.1\mbox{-}1
M_{\odot}/\mbox{yr}$ at $z>6$.  Photometric
redshifts, magnification values, rest-frame properties and
supporting information are all made publicly available as described in
the Appendix~\ref{sec:app}.

\section*{Acknowledgements}
Based on observations made with the NASA/ESA Hubble Space Telescope,
obtained at the Space Telescope Science Institute, which is operated
by the Association of Universities for Research in Astronomy, Inc.,
under NASA contract NAS 5-26555. Observations were also
carried out using {\Spitzer} Space Telescope, which is operated by the
Jet Propulsion Laboratory, California Institute of Technology under a
contract with NASA.  Support for this work was provided by NASA
through ADAP grant 80NSSC18K0945, NSF grant AST 1815458 and AST
1810822 and NASA/HST through {\hst}-AR-14280, {\hst}-AR-13235, {\hst}-GO-13459, {\hst}-GO-13666 and through
an award issued by JPL/Caltech. TT acknowledges support by the Packard
Fellowship.  AH acknowledges support by NASA Headquarters under the
NASA Earth and Space Science Fellowship Program Grant
ASTRO14F-0007. CM acknowledges support provided by NASA through the
NASA Hubble Fellowship grant HST-HF2-51413.001-A.

\appendix
\section{Public release of the Catalogues}
\label{sec:app}
All the catalogues and derived quantities described in this paper are
publicly released and can be obtained from 
 these urls.\footnote{\url{https://doi.org/10.17909/t9-4xvp-7s45}}$^{,}$\footnote{\url{http://www.astrodeep.eu/frontier-fields/}}
Photometric redshift catalogues contain all the photometry as
described in Sect.~\ref{sec:dataanalysis}. These catalogues also contain
photometric redshift properties using EAZY \citep{brammer08}.

Stellar properties  catalogues contain the same information as
catalogues released by \citep{castellano16, dicriscienzo17}. In
particular::
\begin{itemize}
  \item {\tt ID}: identification number that matches the number in the input photometric
    catalogues.
    \item{\tt ZBEST}: corresponds to the reference photo-z value
      used in fitting stellar properties ($z_{\rm best}$). We use
      spectroscopic redshift where available, and photometric
      redshift from EAZY otherwise.  Sources for
which the photo-z run did not converge to a solution or have
unreliable photometry are set
to $ZBEST=-1.0$.
\item {\tt MAGNIF}: median magnification from all the models with
  version 4
  data from this url$^{\ref{fn:lm}}$.
\item {\tt CHI2\_NEB}: $\chi^2$ of the SED fitting with stellar plus nebular templates at
redshift fixed to {\tt ZBEST}.
\item{\tt MSTAR\_NEB, MSTAR\_MIN\_NEB, MSTAR\_MAX\_NEB}: stellar
mass ($10^9 M_{\odot}$) estimated from stellar plus nebular fits.
\item {\tt SFR\_NEB, SFR\_MIN\_NEB, SFR\_MAX\_NEB}: star-formation rate
  ($M_{\odot}/\mbox{yr}$) estimated from the stellar plus nebular fits.
 \item  {\tt CHI2\_NONEB}, {\tt MSTAR\_NONEB, MSTAR\_MIN\_NONEB, MSTAR\_MAX\_NONEB}, {\tt SFR\_NONEB,
     SFR\_MIN\_NONEB, SFR\_MAX\_NONEB} similar to the quantities above, but SED
   fitting was performed using  stellar templates only. Throughout the
   paper we quote all results from SED fitting using stellar plus
   nebular templates, but add these values to the catalog for
   convenience.  
 \end{itemize}
 
\bibliographystyle{mnras}
\interlinepenalty=1000
\bibliography{bibliogr_clusters,bibliogr_gglensing,bibliogr,bibliogr_cv,bibliogr_highz}
\bsp	
\label{lastpage}
\end{document}